\pdfoutput=1
\documentclass[twocolumn,showpacs,twoside,10pt,prl]{revtex4}
\usepackage{graphicx}
\usepackage{epsfig}
\usepackage{epsf}
\usepackage{amssymb}
\usepackage{amsmath}
\usepackage{amsthm}
\usepackage{multirow}
\usepackage{hyperref}

\usepackage[usenames]{color}
\definecolor{Red}{rgb}{1,0,0}
\definecolor{Blue}{rgb}{0,0,1}

\begin{document}

% Include your paper's title here
\title{Quantum Factorization of 143 on a Dipolar-Coupling NMR system}
\author{Nanyang Xu$^{1}$}
\author{Jing Zhu$^{1,2}$}
\author{Dawei Lu$^{1}$}
\author{Xianyi Zhou$^{1}$}
\author{Xinhua Peng$^{1}$}
\email{xhpeng@ustc.edu.cn}
\author{Jiangfeng Du$^{1}$}
\email{djf@ustc.edu.cn}

\affiliation{$^{1}$Hefei National Laboratory for Physical Sciences at
Microscale and Department of Modern Physics, University of Science
and Technology of China, Hefei, Anhui, 230026, China}
\affiliation{$^{2}$Department of Physics $\&$ Shanghai Key Laboratory for
Magnetic Resonance, East China Normal University Shanghai 200062}

\begin{abstract}

Quantum algorithms could be much faster than classical ones in solving the factoring problem.  Adiabatic quantum computation for this is an alternative approach other than Shor's algorithm.
Here we report an improved adiabatic factoring algorithm and its
experimental realization to factor the number $143$ on a liquid
crystal NMR quantum processor with dipole-dipole couplings. We believe this to be the largest number factored in quantum-computation realizations, which shows the practical importance of adiabatic quantum algorithms.

\end{abstract}

\maketitle

Multiplying two integers is often easy while its inverse operation -
decomposing an integer into a product of two unknown factors - is
hard. In fact, no effective methods in classical computers is
available now to factor a large number which is a product of two
prime integers\cite{Knuth}. Based on this lack of factoring ability,
cryptographic techniques such as RSA have ensured the safety of
secure communications\cite{Koblitz}. However, Shor proposed his
famous factoring algorithm\cite{Shor_algorithm} in 1994 which could
factor a larger number in polynomial time with the size of the
number on a quantum computer. Early experimental progresses have been
done to demonstrate the core process of Shor's algorithm on liquid-state
NMR\cite{Chuang} and photonic systems\cite{Luchaoyang,Lanyon} for
the simplest case - the factoring of number 15.

While traditional quantum algorithms including Shor's algorithm are
represented in circuit model, \emph{i.e.}, computation performed
by a sequence of discrete operations, a new kind of quantum
computation based on the adiabatic theory was proposed by Farhi
\emph{et al.} \cite{Farhi} where the system was driven by a
continuously-varying Hamiltonian. Unlike circuit-based quantum algorithms, adiabatic quantum
computation (AQC) is designed for a large class of
optimization problems - problems to find the best one
among all possible assignments. Moreover, AQC shows a better robustness
against error caused by dephasing, environmental noise and
imperfection of unitary operations \cite{Childs_2002,Roland_robust}.
Thus it has grown up rapidly as an attractive field of quantum
computation researches.

Several computational hard problems have been formulated as
optimization problems and solved in the architecture of AQC, for
example the 3-SAT problem, Deutsch's problem and quantum database
search\cite{Farhi,Roland_2002,Das_2002,Xu_search,Steffen,Mitra_2005}.
Recently Peng \emph{et al.} \cite{peng_2008} have adopted a simple
scheme to solve the factoring problem in AQC and implemented it on
a liquid-state NMR system to factor the number 21. However, this scheme
could be very hard for large applications due to the
exponentially-growing spectrum width of the problem Hamiltonian.
%%
%Meanwhile, a more sophisticated scheme to convert the factoring
%problem to an optimization problem had been proposed by
%Burges\cite{Burges} in the classical computer fields. 
At the same time, another adiabatic factoring scheme provided by Schaller and Sch$\ddot{u}$tzhold \cite{Schutzhold,Schaller}
could suppress the spectrum width and shows to be much faster
than classical factoring algorithms or even an exponential
speed-up.

However, Schaller and Sch$\ddot{u}$tzhold's original factoring scheme is too hard to be implemented for any nontrival factoring cases on current quantum processors.  In this letter, we improve the original scheme to use less resources by simplifing the equations mathematically. And a factoring case of  $143$ is choosed as an example to be resolved in this scheme and finally experimentally implemented on a liquid-crystal NMR system with dipolar couplings. We believe this to be the largest number factored on quantum computation realizations.

As mentioned before, AQC was originally  proposed to solve the
optimization problem. Because the solution space of an optimization
problem grows exponentially with the size of problem, to find the
best one is very hard for the classical computers when the problem's
size is large. In the framework of AQC, a quantum system is
prepared in the ground state of initial Hamiltonian $H_0$ , while the
possible solutions of the problem is encoded to the eignestates state of problem Hamiltonian $H_p$
 and the best solution to its ground state. For the
computation, the time-dependent Hamiltonian varies from $H_0$ to $H_p$, and
if this process performs slowly enough, the quantum adiabatic
theorem will ensure the system stays in its instantaneous ground
state. So in the end, the system will be in the ground state of
$H_p$ which denotes the best solution of the problem. Simply the
time-dependent Hamiltonian is realized by an interpolation scheme
\begin{equation}
H(t)=[1-s(t)]H_{0}+s(t)H_{P}\,, \label{e.HSys}
\end{equation}
where the function $s(t)$ varies from 0 to 1 to parametrize the
interpolation. The solution of the optimization problem could
be determined by an measurement of the system after the
computation.

Here, the factoring problem is expressed as a formula $N=p\times q$, where
$N$ is the known product while $p$ and $q$ are the prime factors to be found.
The key part of adiabatic factoring algorithm is to convert the factoring problem to an optimization problem,
and solve it under the AQC architecture. The most straightforward scheme
is to represent the formula as an equation
$N-pq=0$ and form a cost function $f(x,y)=(N - xy)^2$, where $f(x,y)$ is a
non-negative integer and $f(p,q)=0$ is the minimal value of the
function. The problem Hamiltonian $H_p$ could be constructed with the same
form of $f(x,y), $ \emph{i.e.} $H_p=[N - \hat{x}\times\hat{y}]^2$,
where both $\hat{x}$ and $\hat{y}$ are number operators formed by $\sum_{i=0}^{n-1}{2^i\left(\frac{1-\sigma^{i}_z}{2}\right) }$ where $n$ is the bit width of the number and $\sigma^{i}_z$ is the $\sigma_z$ operator representing the $i$th bit. Thus
the ground state of $H_p$ has the zero energy which denotes the case
that $N=xy$. After the adiabatic evolution and measurement, we
could get the result $p$ and $q$. Peng \emph{et al.}
\cite{peng_2008} have implemented this scheme experimentally to factor
21. However in this scheme, the spectrum of problem Hamiltonian
scales with the number $N$, thus it is very hard to implement in
experiment when $N$ is large.

To avoid this drawback, Schaller and Sch$\ddot{u}$tzhold\cite{Schaller} adopted
another scheme by Burges\cite{Burges} to map the factoring problem
to an optimization problem. Their adiabatic factoring algorithm
starts with a binary-multiplication table which is shown in
Tab.\ref{bmt}. In the table, $p_i$ and $q_i$ in the first two rows represent the bits
of the multipliers and the following four rows are the intermediate
results of the multiplication and $z_{ij}$ are the carries from $i$th
bit to the $j$th bit. The last row is the binary representation of number $N$
to be factorized. In order to get a nontrivial case, we set $N$ to
be odd, thus the last bit (\emph{i.e.}  the least significant bit) of
multipliers is binary value $1$. The bit width of $N$ should equal
the summation of the two multipliers' width. For a given product N, the combinations of
these two multipliers' bit width are bounded linearly with $n$. So we
could just focus on a specific combination where the bit widths of the numbers equals with each other and set the first bit
(\emph{i.e.} most significant bit) to be $1$.

%\begin{eqnarray}
%p_1+q_1 &=& 1 + 2z_{12}  \nonumber\\
%p_2+p_1q_1+q_2 +z_{12} &=& 1 + 2z_{23} +4z_{24}\nonumber\\
%1+p_2q_1+p_1q_2+1 +z_{23}&=&1 + 2z_{34} +4z_{35}\nonumber\\
%q_1+p_2q_2+p_1+z_{34}+z_{24}&=& 0 + 2z_{45} +4z_{46}\nonumber\\
%q_2+p_2+z_{45}+z_{35}&=&0+ 2z_{56} +4z_{57}\nonumber\\
%1+z_{56}+z_{46} &=&0+ 2z_{67}\nonumber\\
%z_{67}+z_{57} &=& 1.
%\end{eqnarray}

Then, the factoring equations could be got from each column in
Tab.\ref{bmt}, where all the variables $p_i,q_i,z_{ij}$ in the equations are binary.
To construct the problem Hamiltonian, first we construct bit-wise
Hamiltonian for each equation by directly mapping the binary
variables to operators on qubits, \emph{i.e.} $ H_p^1=
(\hat{p}_1+\hat{q}_1 - 1 -2\hat{z}_{12})^2$, where each symbol with
hat is the corresponding operator $\frac{1-\hat{\sigma}_z}{2}$ of
the qubit for each variable. Then the problem Hamiltonian $H_p= \sum{H_p^{i}}$ is a
summation of all the bit-wise Hamiltonians. In this way, the ground state of $H_p$
encodes the two factors that satisfy all the bit-wise
equations and is the answer to our factoring problem. Thus the
spectrum of $H_p$ will not scale with $N$ but $log_2{N}$.

\begin{table}[ht]
\begin{center}

\begin{tabular}{c|c|c|c|c|c|c|c}

\multicolumn{1}{c}{}

\ & \ & \ & \ & $1$ &  $p_2$  &  $p_1$  &  1  \\
\ & \ & \ & \ & $1$ & $q_2$  &  $q_1$  &  1  \\

\hline
\ & \ & \ & \ & $1$ &  $p_2$  &  $p_1$  &  1  \\
\ & \ & \ & $q_1$ &  $p_2 q_1$  &  $p_1 q_1$  &  $q_1$  & \ \\
\ & \ & $q_2$ &  $p_2q_2$  &  $p_1q_2$  &  $q_2$ & \ & \   \\
\ & $1$ &  $p_2$  &  $p_1$  &  1 & \ & \ & \   \\
\hline
$z_{67}$ & $z_{56}$ & $z_{45}$ &$z_{34}$ &$z_{23}$ &$z_{12}$ & \ &\ \\
$z_{57}$ & $z_{46}$ & $z_{35}$ &$z_{24}$ & \  & \ & \ &\ \\

\hline
$1$ & $0$ & $0$ & $0$ & $1$ & $1$ & $1$ & $1$ \\

\hline
\end{tabular}
\end{center}
\caption{Binary multiplication table. The top two rows are binary representation of
the multipliers whose first and last bit are set too be $1$.
The bottom row is the bits of the number to be factorized which in
our example is $143$. $z_{ij}$ is the carry bit from the $i$th bit
to the $j$th bit in the summation. The significance of each bit in
the column increases from right to left.} \label{bmt}
\end{table}

However, for the example of our interest, the Schaller and
Sch$\ddot{u}$tzhold's scheme\cite{Schaller}  need at least $14$ qubits to factor the
number 143, which exceeds the limitation of current quantum
computation technology. So in our experiment, we introduce
a mathematical simplification to reduce the number of equations in the above scheme due to the requirement that all the variables should be $0$ or $1$.
The first equation $p_1+q_1 = 1 + 2z_{12}$, for example, $z_{12}$ must
be $0$ otherwise the equation could not be satisfied, and so $p_1q_1=0$. After applying the same rule to all the equations,
 we can get a new group of equations, which are : $p_1+q_1=1$, $p_2+q_2=1$ and $p_2q_1+p_1q_2=1$.
Obviously this reduction rule costs a polynomial time with the number of equations.

To construct the problem Hamiltonian from this simplified equations,
the bit-wise Hamiltonians are constructed by
$H^{1}_p=(\hat{p}_1+\hat{q}_1 -1)^2$ , $H^{2}_p=(\hat{p}_2+\hat{q}_2
-1)^2$ and $H^{3}_p=(\hat{p}_2\hat{q}_1+\hat{p}_1\hat{q}_2 - 1 )^2$. But this construction
method causes $H^{3}_p$ to have a four-body interactions, which
is hard to be implemented experimentally. In this case, Schaller and
Sch$\ddot{u}$tzhold\cite{Schaller}  introduced another construction form that for
the equation like $AB+S=0$, the problem Hamiltonian could be
constructed by
$2[\frac{1}{2}(\hat{A}+\hat{B}-\frac{1}{2})+\hat{S}]^2-\frac{1}{8}$,
which could reduce one order of the many-body interactions in
experiment. Thus we replace the third bit-wise Hamiltonian as
$H_p^{'3}=2 [\frac{1}{2} (\hat{p}_1 + \hat{q}_2 - \frac{1}{2}) +
\hat{p}_2\hat{q}_1 - 1]^2 - \frac{1}{8}$. So the problem
Hamiltonian is ,
\begin{eqnarray}
H_p &=& 5-3 \hat{p}_1-\hat{p}_2-\hat{q}_1+2 \hat{p}_1 \hat{q}_1-3 \hat{p}_2 \hat{q}_1\nonumber\\
&+&2 \hat{p}_1 \hat{p}_2 \hat{q}_1-3 \hat{q}_2+\hat{p}_1
\hat{q}_2+2\hat{p}_2 \hat{q}_2+2 \hat{p}_2 \hat{q}_1
\hat{q}_2\nonumber,\label{hp}
\end{eqnarray}
where the operators $\hat{p}$ and $\hat{q}$ are
mapped into the qubits' space as $\hat{p}_1=\frac{1-\sigma_z^1}{2},
\hat{p}_2=\frac{1-\sigma_z^2}{2}, \hat{q}_1=\frac{1-\sigma_z^3}{2}$
and $ \hat{q}_2=\frac{1-\sigma_z^4}{2}. $

For the adiabatic evolution, without the loss of generality, we
choose the initial Hamiltonian $H_0=g(\sigma_{x}^{1}+\sigma_{x}^{2}+\cdots+\sigma_{x}^{n})$ where
$g$ is a parameter to scale the spectrum of $H_0$. And the ground
state of the operator is $|\psi_i\rangle=\left(
\frac{|0\rangle-|1\rangle}{\sqrt{2}} \right)^{\otimes n}$- a
superposition of all the possible states. So for the computation, we
prepare the system on the state $|\psi_i\rangle$ with the Hamiltonian being
$H_0$, and slowly vary the Hamiltonian from $H_0$ to $H_p$ according
to Eq.(\ref{e.HSys}), the quantum adiabatic theorem ensures that the
system will be at the ground state of $H_p$, which represents the
answer to the problem of interests.

\begin{figure}[htb]
\begin{center}
\includegraphics[width= 0.9\columnwidth]{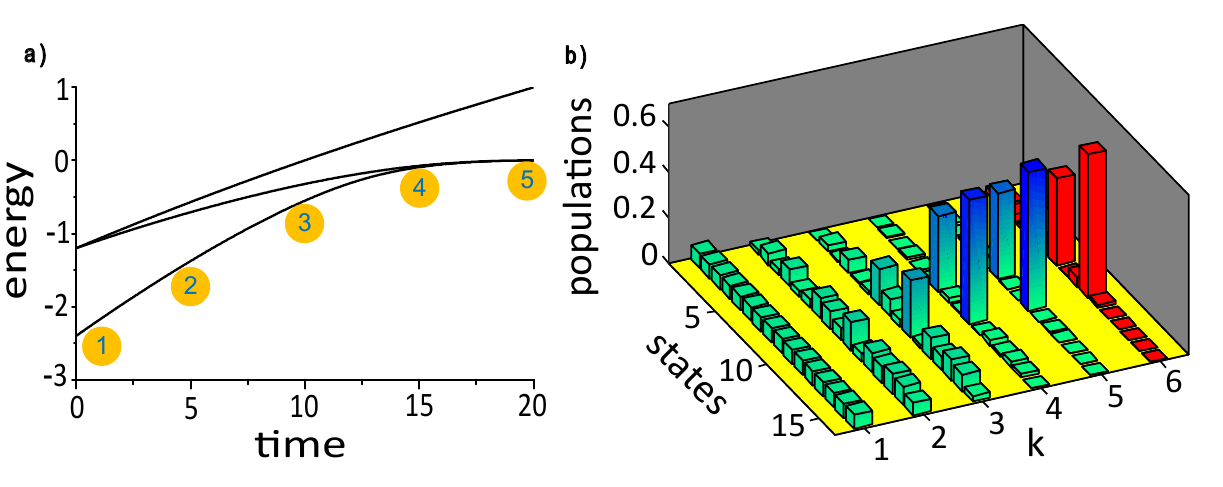}
\end{center}
\caption{Process of the adiabatic factorization of 143. a) the
lowest three energy levels of the time-dependent Hamiltonian in
Eq.(\ref{e.HSys}), The parameter $g$ in the initial Hamiltonian is
$0.6$. b) k=1$\sim$5 shows the populations on computational basis of
the system during the adiabatic evolution at different times marked
in a); k=6 shows the result got from our experiment. The
experimental result agrees well with the theoretical expectation.
The system finally stays on a superposition of $|6\rangle$ and
$|9\rangle$, which denotes that the answer is $\{p=11,q=13\}$ or
$\{p=13,q=11\}$}\label{sim}
\end{figure}

We numerically simulate the process of factoring 143 as shown in
Fig.\ref{sim}. Specially, the ground state of the problem
Hamiltonian in Eq.(\ref{hp}) is degenerated. This is because two
multipliers $p$ and $q$ have the same bit width, thus an exchange of
$p$ and $q$ also denotes the right answer. From the simulation, we
could see that the prime factors of 143 is 11 and 13.

%================================experimental part begin=====================================

\begin{figure}[htbp]
\begin{center}
\includegraphics[width=0.9\columnwidth]{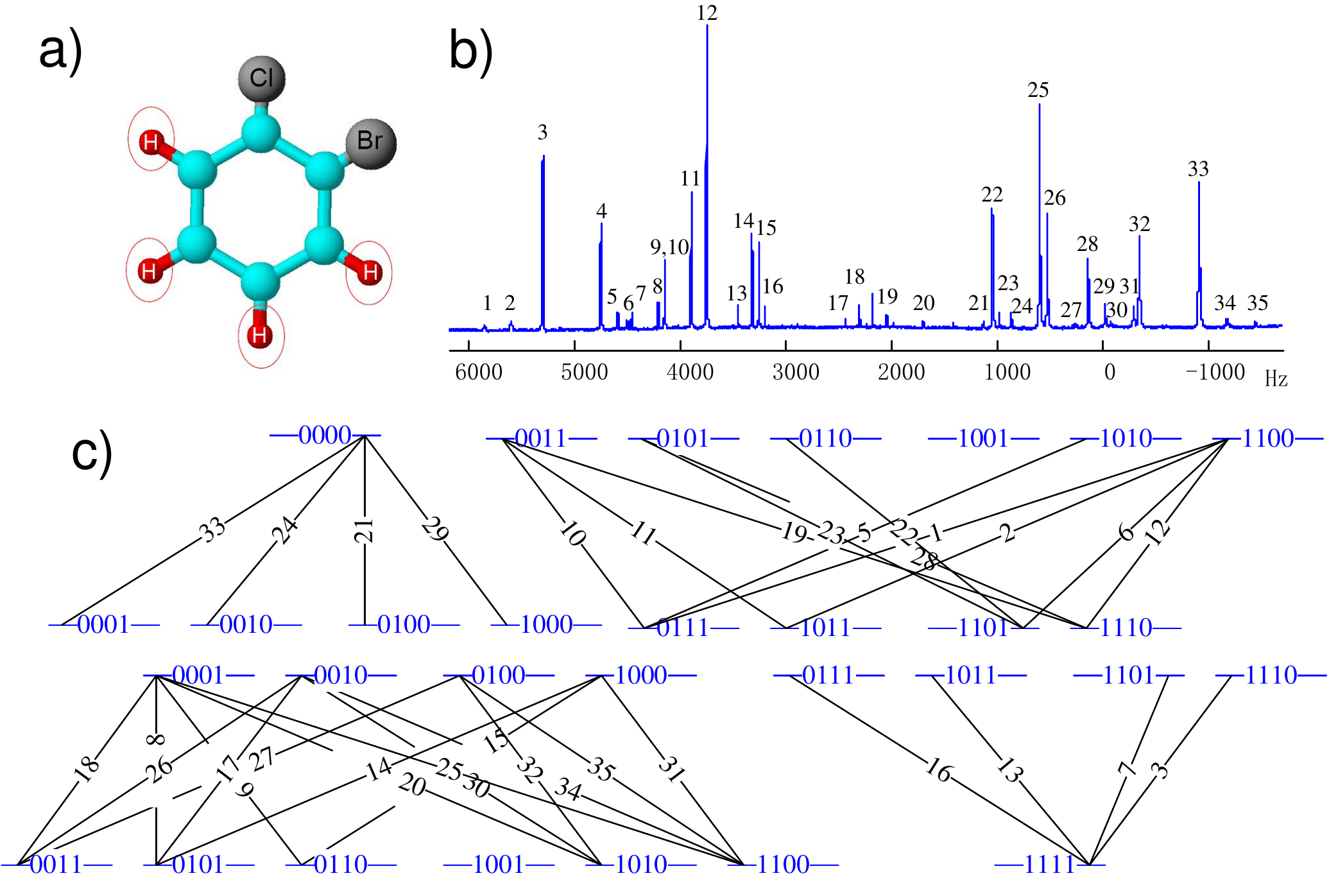}
\caption{\label{sample_spectrum} (Color online) Quantum register in our experiment. a)The
structure of the 1-Bromo-2-Chlorobenzene molecule. The four $^1H$
nuclei in ovals forms the qubits in our experiment. b)
 Spectrum of $^1$H of the thermal state
$\rho_{th}=\sum_{i=1}^4 \sigma_z^i$ applying a $[\pi/2]_y$ pulse. Transitions are labeled
according to descending order of their frequencies. c)Labeling scheme for the states of the
four-qubit system}
\end{center}

\end{figure}
Now we turn to our NMR quantum processor to realize the above
scheme of factoring 143. The four qubits are represented by the four
$^1$H nuclear spins in 1-Bromo-2-Chlorobenzene (C$_6$H$_4$ClBr)
which is dissolved in the liquid crystal solvent ZLI-1132 at
temperature 300 K. The
structure of the molecule is shown in Fig.\ref{sample_spectrum}a and
the four qubits are marked by the ovals. By fitting the thermal
equilibrium spectrum in Fig.\ref{sample_spectrum}b, the natural
Hamltonian of the four-qubit system in the rotating frame is
\begin{eqnarray}
 &\mathcal{H} = 2\pi\sum_i \nu_i
 I_z^i+2\pi\sum_{i,j,i<j}J_{ij}I_z^iI_z^j \nonumber\\
 &+2\pi\sum_{i,j,i<j}D_{ij}(2I_z^iI_z^j-I_x^iI_x^j-I_y^iI_y^j),
\end{eqnarray}
where the chemical shifts $\nu_1=2264.8$Hz, $\nu_2=2190.4$Hz,
$\nu_3=2127.3$Hz, $\nu_4=2113.5$Hz, the dipolar couplings strengthes
$D_{12}=-706.6 $Hz, $D_{13}=-214.0$Hz, $D_{14}= -1166.5$Hz,
$D_{23}=-1553.8$Hz, $D_{24}=-149.8$Hz, $D_{34}=  -95.5$Hz and the
J-couplings $J_{12}=0$Hz, $J_{13}=1.4$Hz, $J_{14}=8$Hz,
$J_{23}=8$Hz, $J_{24}=1.4$Hz, $J_{34}=8$Hz . The labeling transition
scheme for the energy levels is shown in Fig.\ref{sample_spectrum}c.

\begin{figure}[htbp]
\begin{center}
\includegraphics[width=0.8\columnwidth]{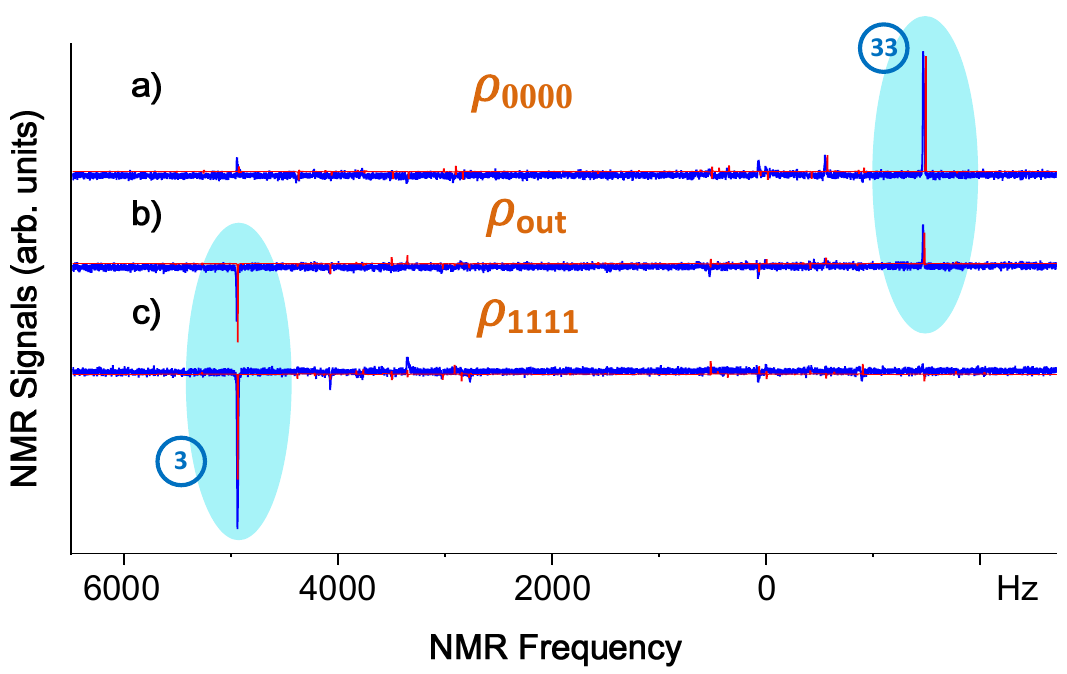}
\caption{\label{lc4r}(Color online) NMR spectra for the small-angle-flip observation of the
PPS and the output state $\rho _{out}$, respectively. The blue spectra (thick) are the experimental results, and the red spectra (thin) are the simulated ones. a) c) Spectra corresponding to the
PPS $\rho _{0000}$ and $\rho _{1111}$ by applying a
small angle flip (3$^{\circ}$) pulse. The main peaks are No. 33 and No. 3 labeled in the thermal equilibrium spectrum. b) Spectrum corresponding to the
output state $\rho _{out}$ after applying a small angle flip
(3$^{\circ}$) pulse, which just consists of the peaks of No. 33 and No. 3.}
\end{center}
\end{figure}

The whole experimental procedure can be
described as three steps: preparation of the ground state of $H_0$, adiabatic passage by the time-dependent Hamiltonian
$H(t)$, and measurement of the final state. Starting from thermal
equilibrium, we firstly created the pseudo-pure state(PPS) $\rho
_{0000}=\frac{1-\epsilon }{16}\mathbb{I}+\epsilon |0000\rangle
\langle 0000|$, where
$\epsilon$ describes the thermal polarization of the system and
${\mathbb{{I}}}$ is an unit matrix. The PPS was prepared from the thermal
equilibrium state by applying one shape pulse based on GRadient Ascent Pulse Engineering (GRAPE) algorithm\cite{Khaneja} and one \emph{z}-direction gradient pulse, with the fidelity 99\% in the numerical simulation.
Fig.\ref{lc4r}a shows the NMR spectrum after a small angle flip
pulse\cite{smallangle} of state $\rho _{0000}$. Then one $\frac{\pi}{2}$ hard pulse was applied  to $\rho
_{0000}$ on the \emph{y}-axis to obtain the ground state of
$H_0$, \emph{i.e.}, $|-\rangle^{\otimes4}$
($|-\rangle=(|0\rangle-|1\rangle)/\sqrt{2}$).

In the experiment, the adiabatic evolution was
approximated by $M$ discrete
steps\cite{Mitra_2005,Steffen,Peng_2005,peng_2008}. We utilized the
linear interpolation $s(t)=t/T$, where $T$ is the total evolution
time. Thus the time evolution for each adiabatic step is
$U_{m}=e^{-iH_m\tau }$ where $\tau=T/M$ is the duration of each
step, and $H_m=( 1- \frac{m}{M} ) H_0 + (\frac{m}{M})H_p$ is the intermediate Hamiltonian of the \emph{m}th step. And the total evolution applied on the initial state is
$U_{ad}=\prod_{m=1}^{M}U_{m}$. The adiabatic condition is satisfied
when $T,M\rightarrow\infty$. Here we chose the parameters $g=0.6,
M=20$ and $T=20$. Numerical simlation shows that the probabilities of
the system on the ground states of $H_p$ is $98.9\%$, which means
that we could achieve the right answer to the factoring problem of
143 almost definitely. We packed together the unitary operators every
five adiabatic steps in one shaped pulse calculated by the GRAPE method\cite{Khaneja}, with the length of each pulse $15ms$ and the fidelity with the theoretical operator over $99\%$. So the total evolution time is about $T_{tot}=60ms$.

Finally we measured all the diagonal elements of the final density
matrix $\rho_{fin}$ using the Hamiltonian's diagonalization method\cite{Dawei}. 32 reading-out GRAPE pulses for population measurement were used after the adiabatic evolution, with
each pulse's length $20ms$. Combined with the normalization condition
$\sum_{i=1}^{16}P(i)=1$, we reconstructed all the diagonal elements of
the final state $\rho_{fin}$. Step $k=6$ of Fig.\ref{sim}b shows the
experimental result of all the diagonal elements excluding the
decoherence through compensating the attenuation factor $e^{-T_{tot}/T_2^{*}}$, where $T_{tot}$ is the total evolution time $60ms$ and $T_2^{*}$ is the decoherence time $102ms$. The experiment (step $k=6$) agrees well with the theoretical expectations (step $k=5$), showing that the factors of 143 is $11$ and $13$.

On the other hand, to illustrate the result more directly from the NMR experiment, a comprehensible spectrum was also given by applying a small angle
flip (3$^{\circ}$) after two $\pi$ operators on the second and third
qubit and one gradient pulse,
\begin{eqnarray}
\rho_{out} = Gz(R_y^{2,3}(\pi)\rho_{fin}R_y^{2,3}(\pi)^{\dagger})
\end{eqnarray}
For the liquid crystal sample, since the Hamiltonian includes non-diagonal elements, the eigenstates are
not Zeeman product states but their linear combinations, except $\left\vert 0000
\right\rangle$ and $\left\vert 1111
\right\rangle$. If there just exist two populations $\left\vert 0000
\right\rangle \left\langle 0000 \right\vert$ and $\left\vert 1111
\right\rangle \left\langle 1111 \right\vert$, the spectrum would be comprehensible as containing only two main peaks after a small angle pulse excitation.  The motivation of adding the $\pi$
pulses after the adiabatic evolution is conversing $\left\vert 0110
\right\rangle \left\langle 0110 \right\vert$ and $\left\vert 1001
\right\rangle \left\langle 1001 \right\vert$ to $\left\vert 0000
\right\rangle \left\langle 0000 \right\vert$ and $\left\vert 1111
\right\rangle \left\langle 1111 \right\vert$, while the gradient
pulse was used to make the output $\rho_{out}$ concentrated on the
diagonal elements of the density matrix. Thus the small angle flip
observation would be easily compared with $\rho_{0000}$ and
$\rho_{1111}$ (Fig.\ref{lc4r}), indicating that the factors of 143 is $11$ and $13$.

%===================================experimental part end====================================

To be concluded, we improved the adiabatic factoring scheme and
implemented it to factor 143 in our NMR platform. The sample we used
for experiment is oriented in the liquid crystal thus it has dipole-dipole
coupling interactions which are utilized for the computation. The
experimental result matches well with theoretical expectations. To
our knowledge, this is the first experimental realization of quantum
algorithms to factor a number larger than 100.

\section{Acknowledgement}

The authors thank Dieter Suter for helpful discussions. This work was supported by National Nature Science Foundation of China (Grants Nos. 10834005, 91021005, and 21073171), the CAS, and the National Fundamental Research Program 2007CB925200.

\end{document}